\documentclass[twocolumn]{article}
          \begin{document}
          %\draft
           \title{Efficient Implementation and the Product State Representation of Numbers}
          \author{Paul Benioff\\
           Physics Division, Argonne National Laboratory \\
           Argonne, IL 60439 \\
           e-mail: pbenioff@anl.gov}
           \date{\today}

          \maketitle
          \begin{abstract}
          The relation between the requirement of efficient
          implementability and the product state
          representation of numbers is examined. Numbers are
          defined to be any model of the axioms of number theory or
          arithmetic.  Efficient implementability (EI)
          means that the basic arithmetic operations are physically
          implementable and the space-time and thermodynamic
          resources needed to carry out the implementations are
          polynomial in the range of numbers considered.  Different
          models of numbers are described to show the independence of
          both EI and the product state representation from the
          axioms. The relation between EI and the product state
          representation is examined.  It is seen that the condition of
          a product state representation does not imply EI.
          Arguments used to refute the converse implication, EI implies a product
          state representation, seem reasonable; but they are not
          conclusive. Thus this implication remains an open
          question.
          \end{abstract}
          %\pacs{03.67.-a,03.65.Ta,03.67.Lx}

\section{Introduction}

In all physical representations of numbers constructed to date,
numbers are represented by strings of numerals or by tensor
product states of systems in quantum mechanics. This is the case
for macroscopic systems, such as classical computers, which are
in such wide use.  It is also true for microscopic systems or
quantum computers which are of much recent interest
\cite{Shor,Grover}.

The universal use of these representations brings up the question,
are these string or tensor product state representations
necessary? Or is it just a matter of convenience rather than
necessity that representations constructed to date have this
property?  This question will be examined here by studying
physical models of the axioms for number theory. Since these
axioms  are supposed to describe natural numbers (the nonnegative
integers), it follows that any physical model of the axioms is a
physical model of the natural numbers.

Since the (nonlogical) axioms of number theory are referred to
often, it is worth stating them explicitly\footnote{Arithmetic
differs from number theory in that Peano's induction axiom is
included.}. In one form they are \cite{Shoenfield,Smullyan}:
$$\begin{array}{ll}1.\; Sw \neq 0 &
2.\; Sw=Sy\rightarrow w=y  \\ 3.\; w+0=w & 4.\; w+Sy=S(w+y) \\
5.\; w\times 0=0 & 6.\; w\times Sy=(w\times y)+w  \\ 7.\;
\neg(w<0) & 8.\; w<y\bigvee w=y\bigvee y<w
\end{array}$$ $$\begin{array}{c} 9.\; w<Sy\leftrightarrow w<y\bigvee
w=y.
\end{array}$$ Here $\bigvee$ and $\neg$ denote "or" and "not" and
$w,y$ are number variables and $S$ is the successor operation.

The reason for the axiomatic approach is that the axioms give a
well defined way to characterize the numbers.  Any physical
system with states and operators that satisfies the axioms has
states that represent the numbers and operators on the states
that represent the arithmetic operations.  Such a system is
referred to as a (physical) model of the axioms.  This definition
is quite useful in that the axioms characterize the natural
numbers in terms of properties of three basic operations, the
successor $S$, addition $(+)$, and multiplication $(\times)$.
These are referred to here as the basic arithmetic operations.

In recent work \cite{BenRNQM}\cite{BenRNQM1}, physical models of
the axioms for the natural numbers, integers, and rational numbers
were studied. Emphasis was placed on the essential role that the
requirement of efficient implementability of the basic arithmetic
operations plays in any physical model of the axiom systems for
the different types of numbers. This requirement is an essential
component of all computers and in studies of computational
complexity \cite{Papadimitriou}.   This requirement is not
expressed by the axiom systems for the different types of
numbers. However, from the viewpoint of the importance of
developing a comprehensive theory of mathematics and physics
together \cite{BenDTVQM},  such a requirement becomes an
essential condition to be satisfied by any physical model of the
axioms.

The condition of efficient implementability applied to the basic
arithmetic operations means that for each operation there must
exist physical procedures that can actually be implemented and for
which the implementation is efficient. Efficiency means that the
space-time and thermodynamic resources needed for implementation
must be polynomial and not exponential in the number of digits in
the numbers represented \cite{BenRNQM}. An equivalent statement
that avoids the use of string representations is that the
resources required must be polynomial in the logarithm of the
numbers represented and not polynomial in the numbers.

Here the position taken follows that in \cite{BenRNQM} in that
any physical model of natural numbers (and integers and rational
numbers also) must satisfy both the axioms of number theory and
the condition of efficient implementability of the basic
arithmetic operations.  That is, a physical system has states
representing numbers if and only if the states can be prepared
efficiently and there exist dynamics for the basic arithmetic
operations that can be efficiently implemented on the states. No
conditions are placed on the complexity of the system.  It can be
macroscopic or microscopic. For microscopic systems for which
decoherence effects  are important \cite{DiVincenzo},  the
requirement is a minimal limit in that it accepts physical
systems on which the basic operations can be applied without loss
of coherence. However, more complex operations requiring more
resources would be affected significantly  by decoherence.

In this paper the interest is in the relations between the axioms
of number theory, efficient implementability, and the product
state representation of numbers. Of special interest is the
question of whether or not efficicient implementability is a
sufficient condition for the states representing numbers to be
product states.  That is, for all physical systems, does efficient
implementability imply a product state representation?  Or do
there exist physical systems for which the basic arithmetical
operations are efficiently implementable on nonproduct state
representations of the numbers?

The wide existence of computers, macroscopic and microscopic,
that are efficient and are based on the product state
representation of numbers, is not of much help in deciding this
question. Is this a matter of convenience in that there also
exist nonproduct representations for which the arithmetic
operations can be efficiently implemented, or can one prove that
no such representations exist?

These relations are investigated by first exploring in more detail
in the next section the relation between physical models and
efficient implementability. Emphasis is placed on quantum
mechanical systems. Then the description of a model is given in
Section \ref{MM} with no assumptions  made about the structure of
the system states representing numbers. The model is based on a
description of operators for several successor operations instead
of just one and on projection operators. Addition and
multiplication are defined in terms of polynomially many
iterations of these simpler operators.

This and other models are used to examine in Section \ref{PSRN}
the relation between efficient implementability, a product state
representation of numbers, and the axioms of number theory. It is
seen that the axioms of number theory are independent of both the
product state representation and the efficient implementability
requirements in that there are models of the axioms in which
these conditions are true and others in which they are false.

Examination of the relation between efficient implementability and
the product state representation shows that the implication,
efficient implementability implies a product state
representation, is an open question, in spite of arguments
suggesting that it is not valid. The converse implication is
proved to be invalid.

\section{Physical Models and Efficient Implementability}
\label{PMEI}
 One way to show the need for the restriction of physical models
 to those satisfying the efficient implementability condition is to
consider physical models of the axioms of arithmetic that do not
satisfy the requirement.  One model that does not use a product
representation consists of a one dimensional lattice of space
positions with a particle located at any one of the positions. If
one site is chosen to be the origin then the state $\psi_{0}$ for
the particle at the origin represents the number  $0$ and the
state $(U_{S})^{n}\psi_{0} =\psi_{n}$ represents the number $n$.
$U_{S}$ implements $S$ by shifting the particle to an adjoining
site in a fixed direction.

In this model the $S$ operation is clearly efficiently
implementable.  However operations for $+$ and $\times$ are not
efficient since their definitions in terms of $S$ show that
exponentially many iterations of $S$ are required.  This model is
a good illustration of the provable fact that any model in which
$+$ and ${\times}$ are defined in terms of iterations of $S$ is
not efficient.

These arguments also extend to any physical models useing product
states for unary representations of numbers. For these models
implementation of $+$ and $\times$ are not efficient irrespective
of whether $S$ is or is not efficient. For this reason, in what
follows product state representations will refer to binary
representations.  Extension to $k-ary$ representations with $k>2$
is straightforward, except that $k$ cannot be too
large\footnote{Basic physical considerations limit the amount of
information that can be placed in or distinguished in a given
space time volume \cite{Lloyd}.} \cite{BenRNQM,Lloyd}.

There also exist physical models with binary product state
representations of numbers in which neither $S,+$ nor $\times$
can be efficiently implemented.  An example consists of a row of
infinite square wells each containing one spinless particle. The
product states representing numbers describe each of the
particles in either the ground or first excited state in the
wells. The wells are scaled so that the well width $d_{j+1}$ for
the well at site $j+1$ is related to that for the site $j$ well by
$d_{j+1}=d_{j}/2$. Since energy level separations in in the $jth$
well are proportional to $(d_{j})^{-2},$ one sees that the energy
resources required to implement any of the basic arithmetic
operations have an exponential dependence on the number $n$ of
wells in the model.

This example shows that the requirement of efficiency can be
separated from that of physical implementability, but only over a
restricted range of physical parameters. For instance for $n\sim
10-15$, such a model could probably be constructed even though it
would not be practical. However for $n\sim 100$ such a model is
impossible to construct as one could not even physically
construct the wells to hold the particles.  This follows from the
scaling of the well size as inversely proportional to the spring
constant. For instance, in this case, if $ d_{1}\sim 1$ cm, then
$d_{100}\sim 10^{-30}$ cm which is of the order of the Planck
length.

Other models to consider represent numbers using entangled
states.  As an example, consider a system of $n$ spin $1/2$
particles contained in potential wells, one particle per well at
positions $x_{1},x_{2},\cdots ,x_{n}$. These are collectively
represented by  a function $ \underline{x}$ from $1,\cdots ,n$ to
the set of $n$ positions. A magnetic field is present as a
reference frame for spin alignment along $(\uparrow )$ and
opposite $(\downarrow )$ to the field direction. Let $
|\underline{s},\underline{x}\rangle
=\otimes_{j=1}^{n}|\underline{s}(j),x_{j}\rangle$ denote a product
state of the spins where $ \underline{s}$ is a function from
$1,\cdots ,n$  to $0,1.$ Here $1$ denotes
$\uparrow$ or spin along and $0$ denotes $\downarrow$ or spin
opposite to the magnetic field direction.

In the following let $ \underline{s}$ be any function as defined
above except that $ \underline{s}(n)=0$ and let $ \overline{s}$
be obtained from $ \underline{s}$ by exchanging ones and zeros at
each location. That is $ \overline{s}(j) = 1 -\underline{s}(j).$
Let $|\underline{s}\rangle$ and $|\overline{s}\rangle$ be the
corresponding product states.  It is clear that all these states
are pairwise orthogonal.

Consider states of the form
$\frac{1}{\sqrt{2}}(|\underline{s},\underline{x}\rangle\pm
|\overline{s},\underline{x}\rangle)$ These entangled states are
also pairwise orthogonal. Numbers can be associated with these
states as follows:
\begin{eqnarray}\frac{1}{\sqrt{2}}(|\underline{s},\underline{x}\rangle +
|\overline{s},\underline{x}\rangle) \Rightarrow
\sum_{j=1}^{n}\underline{t}_{j}2^{j-1}&
\mbox{ if $ \underline{t}=\underline{s}$} \nonumber \\
\frac{1}{\sqrt{2}}(|\underline{s},\underline{x}\rangle
-|\overline{s},\underline{x}\rangle)\Rightarrow
\sum_{j=1}^{n}\underline{t}_{j}2^{j-1} & \mbox{ if $
\underline{t}=\overline{s}$}. \label{entrep}
\end{eqnarray}
It is clear that these states would be difficult, if not
impossible to construct, even in the absence of environmental
decoherence.  Even if they could be constructed, implementation
of the arithmetic operations would be very hard, if not
impossible.  Yet the space resources occupied by these states are
polynomial in $n$ and they are not excluded by the axioms of
number theory.

These examples strongly suggest that the concept of physical
models of the arithmetic axioms should be restricted to models in
which the basic arithmetic operations are efficiently
implementable. In this case one can require that any physical
system of arbitrary complexity has states that represent numbers
(is a physical model of the axioms) if and only if the basic
arithmetic operations are efficiently implementable.  In this
case the states representing numbers are defined by the
properties of the efficient dynamics of the arithmetic operations.

The existence of numerous examples of macroscopic computers, and
hopefully microscopic ones too, that efficiently implement the
arithmetic operations shows that any extension of the axioms of
arithmetic to include efficient implementability would be
consistent. This follows from the fact that an axiom system is
consistent if and only if it has a model \cite{Shoenfield}.
Axiomatization of efficient implementability will not be
attempted here as the concept is still too imprecise. The main
problem is that to say that an operation is implementable means
there exists a physical procedure for carrying out the
operation.  However this requires a precise definition of a
physical procedure which is not yet available.

In spite of this there is much that can be said about this
requirement.  The requirement means that for a given operation
there must exist an efficient implementable dynamics for carrying
out the operation.  In the case of numbers and Schr\"{o}dinger
dynamics, a physical system has states that represent numbers if
and only if there exist Hamiltonians, $H_{S},H_{ +},H_{ \times}$
for efficiently implementing the successor ($S$), addition ($+$),
and multiplication ($\times$) operations on suitable states of the
system.  That is if
$\widetilde{S},\widetilde{+},\widetilde{\times}$ are the
operators on the physical state space of the system that satisfy
the corresponding axioms for number theory, then
\begin{eqnarray}
e^{-iH_{S}t_{S}}\psi \otimes |E\rangle & = &
\widetilde{S}\psi\otimes |E^{\prime}\rangle
\nonumber \\
e^{-iH_{+}t_{+}}\psi \otimes \psi^{\prime} \otimes |E\rangle  & =
& \widetilde{+}\psi\otimes\psi^{\prime}\otimes|E\rangle \nonumber \\
& =&  \psi\otimes (\psi +\psi^{\prime}) \otimes
|E^{\prime}\rangle \label{defplusgen}
\end{eqnarray} and
\begin{eqnarray}
e^{-iH_{\times}t_{\times}}\psi_{\alpha} \otimes \psi_{\beta}
\otimes \psi_{0} \otimes \psi_{0}\otimes|E\rangle  = \\
\widetilde{\times}\psi_{\alpha} \otimes \psi_{\beta} \otimes
\psi_{0} \otimes \psi_{0}\otimes|E\rangle = \\ \psi_{\alpha}
\otimes \psi_{\beta} \otimes \psi_{0} \otimes
(\psi_{\alpha}\times\psi_{\beta})\otimes|E^{\prime}\rangle
\label{deftimesgen}
\end{eqnarray} Here $|E\rangle$ and $|E^{\prime}\rangle$
denote the states of the environment before and after the
interaction. Unitarity requires that the $+$ operation act on
pairs of product states and $\times$ act on quadruples of product
states. The state $\psi_{0} =|0\rangle$ denotes the number $0$.
If each state $\psi$ with a different subscript corresponds to a
linear superposition $\psi =\sum_{j}c_{j}|j\rangle$ where
$|j\rangle$ is the physical state corresponding to the number
$j$, then the dynamics acts  in a standard fashion on each
component $|j\rangle$ in the superposition for $H_{S}$ and on the
product components $|j\rangle\otimes |j^{\prime}\rangle ,$ etc.

Probably the best way to express explicitly the requirement of
efficiency is to note that any dynamical process, such as those
given above for arithmetic operations, is an information
manipulation procedure. Such a process is a sequence of
alternating information acquisition and processing phases. If the
dynamics requires $n$ bits or qubits of information as inputs,
then efficient implementation means that the rate of acquiring and
processing the $n$ bits or qubits must be polynomial and not
exponential in $n$.  This can be expressed crudely as follows:
Let $R_{aq}(t)$ and $R_{pr}(t)$ be the rates, in bits or qubits
per unit time, of information acquisition and processing by some
process.  If these rates are independent of time
then,\begin{equation} R(t) = \left
\{\begin{array}{l}cn^{1-k} \mbox{ if rate polynomial in $n$} \\
cnK^{-n}  \mbox{ if rate exponential in $n$}\end{array}\right.
\label{polyexp}
\end{equation} Here $c,k,K$ are constants that depend on the
dynamics of the process under consideration. They can also be
different for the acquisition and processing phases and any other
relevant system parameters.

If the dynamics of a process require the acquisition and
processing of $n$ bits or qubits of information, the time $t$
required to carry out the process is given approximately by
$\int_{0}^{t}R(t)dt \sim Rt =n$ or $t=c^{-1}n^{k}$ (polynomial)
and $t= c^{-1}K^{n}$ (exponential). Which type applies depends on
both the process dynamics and the state representation used.

If the dynamics of each of the processes for implementing the
three basic arithmetic operations for numbers up to $2^{n}$
requires the acquisition and processing of $n$ bits or qubits,
then efficiency requires that the times $t_{S},t_{+},t_{\times}$
given in Eqs. \ref{defplusgen} and \ref{deftimesgen} are all
equal  to $c^{-1}n^{k}$ where the constants can be different for
each of the three processes.  However it does not follow that for
all physical systems the dynamics of each of these three processes
requires $n$ bits of information.

An example of this is shown by the example considered earlier of
the unary representation of numbers. In this case the successor
operation $S$ is just a shift.  Since implementation of the shift
is independent of where it is, the  information required by the
dynamics is a constant independent of $n$. Stated otherwise the
operation is strictly local.  The dynamics for implementing  $+$
and $\times$ are quite different in that they are global. For
these operations, implementation of the dynamics depends on where
the particle is relative to the choice of the origin, or location
of the $0$ site.  Because of this the dynamics for these two
operations are exponentially slow even though that for $S$ is
 polynomial. This is why unary representations are
rejected as physical models of number theory or arithmetic.

These considerations also show that the condition of efficient
implementation is not preserved under arbitrary unitary
transformations.  If $ \tilde{S}$ is an operator on a Hilbert
space, $\cal H$, of states $\psi$  and $U$ is a unitary operator
acting on $\cal H$, then $U\tilde{S}U^{\dagger}$ acting on
$U\psi$ is equivalent to $\tilde{S}$ acting on $\psi$ in that
$\langle U\psi |U\tilde{S}U^{\dagger}|U\psi\rangle =
\langle\psi|\tilde{S}|\psi\rangle.$  From Eqs. (2)-(4)
(suppressing the environment states) similar equivalences exist
for $ \tilde{+}$ and $\tilde{\times}$. If $W=U\otimes U$ and
$V=W\otimes W$, then $W\tilde{+}W^{\dagger}$ and
$V\tilde{\times}V^{\dagger}$ acting on states $W\Phi$ in ${\cal
H}\otimes {\cal H}$ and $V\Theta$ in ${\cal H}\otimes {\cal
H}\otimes {\cal H}\otimes {\cal H}$ are equivalent to $
\tilde{+}$ and $\tilde{\times}$ acting on $\Phi$ and $\Theta$.
However it does not follow from the efficiency of implementing
$H_{S},H_{+},H_{\times}$ on $\psi ,\Phi , \Theta$ that
$UH_{S}U^{\dagger}, WH_{+}W^{\dagger}, VH_{\times}V^{\dagger}$
are implementable or efficient on $U\psi , W\Phi$, or $V\Theta$.

\section{Multisuccessor Models} \label{MM}

As was noted one problem with defining $+$ and $\times$ in terms
of the successor operation described in the axioms is that
exponentially many iterations are required.  This leads to the
question of finding relatively simple operations whose properties
can be easily axiomatized, and polynomially many iterations of
these operations  can be used to define $+$ and $\times$.

One approach to this problem is to consider a model based on the
use of many successor operations, not just one.  In this model
the $+$ and $\times$ operations are defined in terms of
polynomially many iterations of the successor operations. The
product representation of numbers is not assumed.

The multisuccessor model is motivated  by the binary
representation of numbers $<2^{n}$ shown in the righthand term of
Eq. \ref{entrep}. Based on this representation successor
operators, $S_{1}=S,S_{2},\cdots ,S_{j},\cdots$, are introduced
for each $j$. These operators correspond to addition of $2^{j-1}$
just as $S$ corresponds to the $+1$ operation.

If desired, one may expand the axioms for arithmetic by inclusion
of axioms for all the successor operators.  However this will not
be done here as it is not necessary. Also one may wonder what has
been gained by requiring the efficient implementability of all
the $S_{j}$ rather than applying this requirement separately to
just the three operations $S,+,\times$. One reason is that the
successor operations are simpler operations than are $+$ and
$\times$.  Also in many physical models the $S_{j}$ are related to
one another by means of a  transformation  operator $U$ that is
independent of the index $j$.  That is $S_{j+1}=U(S_{j})$. In
these models, which are much used in practice, efficient
implementability of all the successors follows from that of the
two operations, $S$ and $U.$

The model considered here will be a microscopic model in which
numbers are represented by orthonormal states in a Hilbert space
$\cal H$ with arbitrary tensor product structure.  For example
$\cal H$ could have no tensor product structure or it could be a
tensor product space where the subspaces are described by
different types of entangled states. This includes a possible
description using  bound entangled states as described by Bennett
and others \cite{Bennett}. To keep things simple the model will
be given for arithmetic modulo $2^{n}.$

Let $A$ be a set of physical parameters for a quantum system.
These could be eigenvalues for some system observable. Let
$V_{a}$ be a set of operators  on the state space of the system
indexed by the parameters $a$ in a finite set $A$ of n
parameters. The operators $V_{a}$ are required to have the
following properties \cite{BenRNQM}:
\begin{enumerate}
\item Each $V_{a}$ is a cyclic shift.
\item The $V_{a}$ all commute with one another.
\item There is just one $a$ for which $(V_{a})^{2}=1$. Let $a_{m}$
be this unique $a$.
\item For each $a \neq a_{m}$, there is a unique $a^{\prime}\neq a$ such
that $(V_{a})^{2} = V_{a^{\prime}}$.
\item For each $a^{\prime}$ , if there is an $a\neq a^{\prime}$ such that
$(V_{a})^{2} = V_{a^{\prime}}$, then $a$ is unique.

\item For just one $a$ there are no $a^{\prime}$ such that
$(V_{a^{\prime}})^{2}=V_{a}$. Let $a_{\ell}$ be this unique value.
\end{enumerate}
Properties 3-6 can be used to establish an ordering
$a_{1},a_{2},\cdots ,a_{n}$ of the parameter set $A$ where
$a_{1}=a_{\ell}$ and $a_{n}=a_{m}$ and
$V_{a_{j+1}}=(V_{a_{j}})^{2}$ for $j<n$. Based on this ordering,
the $V_{a_{j}}$ can be considered informally as corresponding to
addition of $2^{j-1}$. The commutativity and cyclic shift
properties give the existence of a set $\cal B$ of pairwise
orthogonal subspaces of states such that for each $a$ and each
subspace $\beta$ in $\cal B$, $V_{a}\beta$ is in $\cal B$ and is
orthogonal to $\beta$.

The properties can also be used to show that there are $2^{n}$
orthogonal subspaces in $\cal B$ that can be given a cyclic
ordering by iterations of $V_{a_{1}}$. However there is no
association of the property parameters in $A$ to the subspaces
$\beta$. Also no subspace is associated with the number $0$. From
now on the subspaces are assumed to be one dimensional, so
$\beta$ can be represented as a state $|\beta\rangle$.

One way to achieve this is  to define operators that can be used
to describe this association.  To this end let $p\epsilon \{\alpha
,\gamma\}$ denote the two values of some physical parameter
associated with an observable that is different from that
associated with the values in $A$. Define $2n$ projection
operators $P_{a,p}$ and $n$ unitary operators $U_{a}$ to have the
following properties:
\begin{enumerate} \setcounter{enumi}{6}
\item Each $P_{a,p}$ is $2^{n-1}$ dimensional and all the
$P_{a,p}$  commute with one another. Also $P_{a,\alpha} =
\widetilde{1}-P_{a,\gamma}$ for each $a$.
\item $U_{a}P_{a^{\prime},p}= P_{a^{\prime},p}U_{a}$ if $a
\neq a^{\prime}$.
\item $U_{a}P_{a,\alpha}= P_{a,\gamma}U_{a};\;\; U_{a}P_{a,\gamma}= P_{a,\alpha}U_{a}.$
\item  For each $a$ there is exactly one $p$ such that
$P_{a,p}|\beta\rangle = |\beta\rangle .$
\end{enumerate}
Properties 7,10 show that to each state $|\beta\rangle$ there is
associated a specific function $ \underline{s}$ from the set $A$
to $\{\alpha,\gamma\}$.  The association is given by
\begin{equation}P_{ \underline{s}}|\beta\rangle =\prod_{a\epsilon
A}P_{a,\underline{s}(a)}|\beta\rangle=|\beta\rangle.\label{sbeta}
\end{equation} Uniqueness is provided by the next property:
\begin{description}
\item[  11.] $P_{\underline{s}}|\beta\rangle
=P_{\underline{s}}|\beta^{\prime}\rangle$ implies that
$|\beta\rangle =|\beta^{\prime}\rangle .$
\end{description} Since there are $2^{n}$ functions $
\underline{s}$ and states $|\beta\rangle$, the above shows that
each $ \underline{s}$ is associated with some $|\beta\rangle$.

The relation of the $P_{a,p}$ and $U_{a}$ to the $V_{a}$ is
provided by the following condition:
\begin{description}
\item[ 12.]   $\begin{array}{lll} V_{a}& =& U_{a}P_{a,\alpha}+V_{Sa}U_{a}P_{a,\gamma}
 \mbox{ if } a\neq
a_{m}; \\ V_{a_{m}}& = &U_{a_{m}}\end{array}$.
\end{description}
Here $a_{m}$ is the value given by property $3$ for the $V_{a}$
and $Sa$ is the unique value of $a^{\prime}$ that satisfies
property $4$. This use of the successor notation is based on the
fact that properties 3-6 of the $V_{a}$ express a successor
operation and an ordering on the set $A$ that satisfies  the
number theory axioms 1,2 and 7-9 listed in the introduction.

These operators can be used to define an addition operator
$\widetilde{+}$ on pairs $|\beta\rangle\otimes
|\beta^{\prime}\rangle$ of states by
\begin{eqnarray}
\widetilde{+}|\beta\rangle\otimes |\beta^{\prime}\rangle &=&
\prod_{a\epsilon A}(P_{a,\alpha}\otimes \widetilde{1}
+P_{a,\gamma}\otimes V_{a})|\beta\rangle\otimes
|\beta^{\prime}\rangle \nonumber \\ &=& |\beta\rangle\otimes|\beta
+ \beta^{\prime}\rangle .\label{defplus}
\end{eqnarray}
The $"+"$ without the tilde in $|\beta + \beta^{\prime}\rangle$
refers to the result of arithmetic addition.  It does not denote
the coherent sum $|\beta\rangle +|\beta^{\prime}\rangle$ of
$|\beta\rangle$ and $|\beta^{\prime}\rangle$.  The unordered
product is used as the operators $ P_{a,\alpha}\otimes
\widetilde{1}+P_{a,\gamma}\otimes V_{a}$ for different $a$
commute with one another.

The unique association of a function $ \underline{s}$ with each
state $|\beta\rangle$, property 10, shows that the addition
operator can also be represented by
\begin{equation} \widetilde{+}|\beta\rangle\otimes |\beta^{\prime}\rangle
=|\beta\rangle\otimes\prod_{a\epsilon A}(V_{a})^{
\underline{s^{\prime}}(a)}|\beta^{\prime}\rangle
.\label{defplus1}\end{equation} Here $ \underline{s^{\prime}}$ is
obtained from $ \underline{s}$ by replacing $\alpha$ with $0$ and
$\gamma$ with $1$.

It follows from the definition of $ \widetilde{+},$ Eq.
\ref{defplus}, that the state $|\beta\rangle$ satisfying $P_{
\underline{\alpha}}|\beta\rangle = |\beta\rangle$ where $
\underline{\alpha}$ is the constant $\alpha$ sequence is the
additive identity.  As shown by the number theory axioms, this
state represents the number $0$.  It follows that any state
$|\beta\rangle$ is related to the $0$ state $|\beta\rangle_{0}$ by
\begin{equation} |\beta\rangle = \prod_{a\epsilon A}(V_{a})^{
\underline{s}(a)}|\beta\rangle_{0}\label{sbeta0}\end{equation}
where $ \underline{s}$ is the unique sequence associated with
$|\beta\rangle$ by Eq. \ref{sbeta}.

To define the multiplication operator it is quite useful to first
define  the operator $W$  by
\begin{equation} W|\beta\rangle =|\beta
+\beta\rangle . \label{defW} \end{equation} $W$ corresponds
informally to the addition of $|\beta\rangle$ to itself.
Iteration of $W$ in Eq. \ref{defW} gives the result that
$W^{j+1}|\beta\rangle=|W^{j}\beta +W^{j}\beta\rangle .$ Use of eq.
\ref{sbeta0}, and Eq. \ref{defplus1} gives the result that
\begin{equation} W^{h+1}|\beta\rangle =
 \prod_{j=1,s_{j}=1}
^{n-h}V_{a_{j+h}}|\beta_{0}\rangle \label{defWh}\end{equation} if
$s_{j}=1$ for some $j\leq n-h$. Otherwise $W^{h+1}|\beta\rangle
=|\beta_{0}\rangle$. It follows that $W^{h}|\beta\rangle
=|\beta_{0}\rangle$ for all $h\geq n+1$.

A definition of $ \widetilde{\times}$ can now be given in terms
of $W$ and $ \widetilde{+}$.  It is defined on triples of states
by \cite{BenRNQM}
\begin{eqnarray}
& \widetilde{\times}|\beta\rangle_{1}\otimes
|\beta^{\prime}\rangle_{2}\otimes|\beta_{0}\rangle_{3} =
\nonumber \\ & \prod_{j=2}^{n}[(P_{a_{j},\alpha}\otimes
\widetilde{1}_{2,3} +P_{a_{j},\gamma}\otimes
\widetilde{+}_{2,3})W_{2}]\times \nonumber \\ &
(P_{a_{1},\alpha}\otimes \widetilde{1}_{2,3}
+P_{a_{1},\gamma}\otimes
\widetilde{+}_{2,3})|\beta\rangle_{1}\otimes
|\beta^{\prime}\rangle_{2}\otimes |\beta_{0}\rangle_{3} \nonumber
\\ & = |\beta\rangle_{1}\otimes|\beta_{0}\rangle_{2}\otimes
|\beta\times\beta^{\prime}\rangle_{3}. \label{deftimes}
\end{eqnarray} Here $W$ is defined by Eq. \ref{defW} and the
subscripts $"2"$ and $"2,3"$ on the operators refer to the state
subscripts in the triple product. $|\beta_{0}\rangle$ represents
the number $0$.

As defined $ \widetilde{\times}$ is not unitary.  This can be
fixed by expanding $ \widetilde{\times}$ to act on quadruples of
the form
$|\beta\rangle_{1}\otimes|\beta^{\prime}\rangle_{2}\otimes
|\beta_{0}\rangle_{3}\otimes |\beta_{0}\rangle_{4}$. One starts
by copying $|\beta^{\prime}\rangle_{2}$ to
$|\beta_{0}\rangle_{4}$. Then at the conclusion of the action,
$|\beta_{0}\rangle_{2}$ and $|\beta^{\prime}\rangle_{4}$ are
exchanged. Also in order to ensure unitarity $
\widetilde{\times}$ was defined to add the result of
multiplication to whatever state is the $3rd$ component.  That is,
if $|\beta^{\prime\prime}\rangle_{3} \neq |\beta_{0}\rangle_{3}$,
the final $3rd$ state component can be represented as
$|\beta^{\prime\prime}+(\beta\times\beta^{\prime})\rangle.$

\section{Is the Product State Representation
Necessary?}\label{PSRN}

There is much to discuss about the results obtained so far. One
feature is that each state $|\beta\rangle$ is in a simultaneous
eigenstate of all the values $a$ in $A.$ This follows from
property 10. If $q_{a}$ is the projection operator for an
eigenspace associated with $a$ then $q_{a}|\beta\rangle =
|\beta\rangle$ for all $a$ and all $|\beta\rangle$.

This may seem counterintuitive but this property is satisfied by
most product state models. For example, let $A$ be a set of $n$
space positions of potential wells each containing a single spin
$1/2$ particle. There is a common magnetic field to determine the
spin direction.  Product states have the form
$|\underline{s},A\rangle = \otimes_{a\epsilon
A}|\underline{s}(a),a\rangle$, or
$|\underline{s},\underline{a}\rangle =
\otimes_{j=1}^{n}|\underline{s}(j),\underline{a}(j)\rangle$ in a
more standard form.  In the second form  $\underline{s}$ and $
\underline{a}$ are respective functions from $1,\cdots ,n$ to
$\{\uparrow ,\downarrow\}$ and from $1,\cdots ,n$ to $A$.  It is
clear that for any of the $2^{n}$ states
$|\underline{s},\underline{a}\rangle ,$
$q_{a}|\underline{s},\underline{a}\rangle
=|\underline{s},\underline{a}\rangle$ for each $a\epsilon A$ and
all $ \underline{s}$.

Based on this one might conclude that the properties of the
$V_{a}$ and the projection and unitary operators given above are
sufficient to prove that the states $|\beta\rangle$ have a product
structure. This is not the case.

To see this consider the entangled state representation of
numbers by Eq. \ref{entrep} for the model described above. In
this model let $Q_{ \underline{s},A}$ and $Q_{ \overline{s},A}$ be
projection operators for the states $ |\underline{s},A\rangle$
and $ |\overline{s},A\rangle$ respectively. That is, $Q_{
\underline{s},A} |\underline{s},A\rangle =
|\underline{s},A\rangle$ and $Q_{ \overline{s},A}
|\overline{s},A\rangle = |\overline{s},A\rangle. $ Here, as
before, $ \underline{s}(a_{m}) =0$ and  $ \overline{s}(a)
=1-\underline{s}(a)$ for each $a\epsilon A$.  $a_{m}$ is the
maximum value of $A$ according to property 3.

Define the unitary operator $U$ by
\begin{eqnarray} \frac{1}{\sqrt{2}}(|\underline{s},A\rangle +
|\overline{s},A\rangle)=U|\underline{s},A\rangle  \nonumber \\
\frac{1}{\sqrt{2}}(|\underline{s},A\rangle -
|\overline{s},A\rangle)=U|\overline{s},A\rangle . \label{defU}
\end{eqnarray}  Unitarity follows
from the fact that
$\langle\underline{s^{\prime}},A|U^{\dagger}U|\underline{s},A\rangle
=\delta_{  \underline{s},\underline{s^{\prime}}}.$

Based on this one sees that $ P_{ \underline{t},A} = UQ_{
\underline{t},A}U^{\dagger} $ for any sequence $ \underline{t}$
where $ \underline{t} = \underline{s}$ or $ \underline{t} =
\overline{s}.$ So $P_{\underline{t},A}$ satisfies Eq. \ref{sbeta}
with $|\beta\rangle = U|\underline{t},A\rangle$.  Note that in
Eq. \ref{sbeta} $P_{ \underline{s}}\equiv P_{ \underline{s,A}}.$

From this one has $$P_{a,p}=\sum_{ \underline{t}}^{
\underline{t}(a)=p}P_{\underline{t},A}.$$ If $U_{a}$ and $V_{a}$
are defined by properties 7-12, it is straightforward to show
that the $V_{a}$ have properties 1-6. In this case the definitions
of $ \widetilde{+}$ and $ \widetilde{\times}$ in terms of these
operators apply.  Proofs that these operators satisfy the axioms
of number theory are tedious but also straight forward
\cite{BenRNQM,BenRNQM1}.

This constitutes a proof that nothing in the axioms of number
theory implies a product state representation model, even for
multiple successor models based on the projection operators and
the $V_{a}$ with the properties described.  It follows that the
axioms of number theory are independent of the product state
representation condition in that there are models of the axioms
in which numbers are represented by product states and models in
which they are represented by entangled states.

The number theory axioms are also independent of the requirement
that the basic arithmetic operations are efficiently
implementable. This is shown by  both the  well known existence
of physical models in which the operations are physically
implementable  and the example given in Section \ref{PMEI} of a
model containing a row of infinite square wells wells where the
well width decreased exponentially with well position. For this
example the operations are not efficient and are therefore not
efficiently implementable.

It remains to address the relation between the requirement of
efficient implementability and product state representations of
numbers.  The example noted above shows that the implication:
product state representation of numbers implies the efficient
implementability of the basic arithmetic operations is not valid.
The reverse implication is more difficult.  In fact one can give
arguments that suggest that efficient implementability is
independent of the product representation of numbers. That is, it
neither implies or is implied by the product representation
condition.

It is worth examining this in more detail. To prove that
efficient implementability does not imply a product state
representation it is sufficient to show some entangled
representation, such as that for Eq. \ref{entrep}, for which the
successor operators $V_{a}$ defined by properties $1-6,12$ are
efficiently implementable.

To this end assume the entangled representation of numbers given
by Eq. \ref{entrep} with the $n$ physical systems located as
described at space sites $x_{1},\cdots ,x_{n}$.   Then the
physical procedure for implementing each $V_{a}$ would have to
include coherent interactions with all the $n$ physical systems.
The interactions between the component systems would have to
extend coherently over the space region occupied by the $n$
systems.

It is reasonable to expect that the degree of difficulty, or
resources needed, to implement the $V_{a}$ would increase
polynomially with $n$.  This is based on the argument that the
range over which the interactions need to be coherent increases
linearly with $n$. This suggests that if the $V_{a}$ are
efficiently implementable for physical states of the form of Eq.
\ref{entrep} for some $n$, they are efficiently implementable for
all $n$ even though the resources required for implementation
might increase with a high power of $n$. One would not expect the
resources required to increase exponentially with $n$.

This type of inductive reasoning, combined with the fact that for
$n =2$ the two operators $V_{a}$ should be physically
implementable, suggests that the implication is valid. Physical
implementability for $n=2$ is based on the the fact that the
states shown in Eq. \ref{entrep} are  the four Bell states.

The problem with this argument is that, although it may be
reasonable, it does not constitute a rigorous proof. Lacking is a
discussion of the $n$ dependence of the resources required to
overcome the effects of decoherence \cite{Zurek,Zeh} including
the use of quantum error correction codes \cite{QEC}. Also
lacking is a precise definition of physical implementability of a
procedure. Without this it is difficult to show conclusively, in
spite of the above argument of reasonableness, that efficient
implementability does not imply a product representation of
numbers.

The above shows that the properties of numbers and the basic
arithmetic operations cannot be used to determine if efficient
implementability implies a product state representation of
numbers. One must look elsewhere for such a proof. Another
approach is based on the fact that all physical processes and
computations are specific examples of information manipulation
processes.  In general each such process consists of a sequence
of alternating information acquisition phases, information
processing  phases, and possible information distribution
phases.  This includes computations and tasks performed by
robots, microscopic \cite{BenQR} or macroscopic.

If the dynamics of an information manipulation process depends on
or is sensitive to $n$ bits or qubits of information then at least
$n$ bits or qubits of information must be acquired, and
processed.  Then the  (reversible) dynamics of the process is
represented by a unitary step operator $U$ acting on the $2^{n}$
dimensional Hilbert space of states of the $n$ qubits.  Since one
is interested in the time development of the states of the $n$
qubits, it makes sense to choose the product basis
$|\underline{b}\rangle
=\otimes_{j=1}g^{n}|\underline{b}_{j}\rangle_{j}$, where
$|\underline{b}_{j}\rangle_{j}$ is a basis state for the $jth$
qubit, as the reference basis for the $n$ qubits rather than some
entangled basis.

This abstract representation of the dynamics of an $n$ qubit
information theoretic process is related to physical processes
through unitary maps $W$ from the basis states
$|\underline{b}\rangle$ to a  basis of physical states of some
physical system that span a $2^{n}$ dimensional Hilbert subspace
of states of the system \cite{BenRNQM}. (See also Viola et al
\cite{Viola} for a discussion regarding the relation between
qubits and physical systems.)  The dynamical process on the states
of the physical system corresponding to the action of $U$ on the
qubits is represented by the operator $WUW^{\dagger}$.

It is to be noted that there is no requirement that the map $W$
take product qubit states into product states of different
physical degrees of freedom of the physical system.  The states
$W|\underline{b}\rangle$ can just as well be entangled states of
the physical system. Whether they are entangled or product states
depends on $W$.

It is also the case that the requirement of efficient
implementability applies to the implementation of the operator
$WUW^{\dagger}$ as this corresponds to a physical process.  The
requirement does not apply to the more abstract $U$ as this is an
abstract information theoretic dynamics representing many
different physical processes, each characterized by a different
map $W$ from the information theoretic qubit states to different
Hilbert spaces of physical states of different systems.

This situation makes it unlikely that anything is to be gained by
using the more abstract information dynamics to prove or disprove
that efficient implementability implies or does not imply a
product state representation.  If one could prove the
implication, then this would restrict the maps $W$ to be maps
from product qubit states to product states of physical degrees
of freedom.  One must conclude that the implication, efficient
implementability of a process implies a product state
representation of the physical states of a system on which the
process is to be carried out, is an open question.

\section{Discussion}

It must be emphasized that the arguments given before to suggest
that the implication does not hold for states representing
numbers do not constitute a proof. As such they do not contradict
the open question conclusion stated above.  As has been noted a
problem in giving such a proof is the lack of an exact
characterization of physical implementability.  Lacking this, it
is difficult to make further progress in this direction.

However, the work done here does show that the conditions of
efficient implementability and of a product state representation
of numbers are independent of the axioms of number theory. The
result that information theoretic arguments do not help to
detemine the validity of the implication, efficient
implementability implies product state representation,  is a
consequence of the assumed separation of abstract qubit states
and their dynamics from states and dynamics of real physical
processes to which they are related through the maps $W$. If this
assumed picture turns out not to be valid, then the argument may
have to be revised.

\section*{Acknowledgements}
This work is supported by the U.S. Department of Energy, Nuclear
Physics Division, under contract W-31-109-ENG-38.

        \end{document}